\documentclass[aps,twocolumn,superscriptaddress,showpacs,amssymb,prl,floatfix,preprintnumbers]{revtex4-1}
\usepackage{graphicx,color}
\usepackage{dcolumn}
\usepackage{amsmath}
\usepackage{amssymb}
\usepackage{epstopdf} 
\usepackage{dsfont}
\usepackage[dvipsnames]{xcolor}
\usepackage{ulem}

\usepackage{hyperref}
\hypersetup{colorlinks=true,linkcolor=blue,pagebackref=true,
  implicit=true,breaklinks=true,pagebackref=true,backref=true,
  bookmarks=true,bookmarksnumbered=true,hyperfootnotes=true,debug=true,
  naturalnames=false,citecolor=blue,pdfview=FitH,pdfstartview=FitH,hyperindex=true}

\newcommand{\CeCoIn}     {CeCoIn$_5$}
\newcommand{\CeRhIn}     {CeRhIn$_5$}
\newcommand{\CeIrIn}    {CeIrIn$_5$}
\newcommand{\pd}{{\phantom{\dagger}}}

\begin{document}

\thispagestyle{myheadings}

\title{Dirac fermions in the heavy-fermion superconductors Ce(Co,Rh,Ir)In$_5$} 

\author{Kent R. Shirer}
\email{kent.shirer@cpfs.mpg.de, yan.sun@cpfs.mpg.de}
\thanks{These authors contributed equally to this work.}
\author{Yan Sun}
\email{kent.shirer@cpfs.mpg.de, yan.sun@cpfs.mpg.de}
\thanks{These authors contributed equally to this work.}
\affiliation{Max Planck Institute for Chemical Physics of Solids, 01187 Dresden, Germany}
\author{Maja D. Bachmann}
\affiliation{Max Planck Institute for Chemical Physics of Solids, 01187 Dresden, Germany}
\author{Carsten Putzke}
\affiliation{Institute of Materials, \'{E}cole Polytechnique F\'{e}d\'{e}rale de Lausanne (EPFL), 1015 Lausanne, Switzerland}
\affiliation{Max Planck Institute for Chemical Physics of Solids, 01187 Dresden, Germany}
\author{Toni Helm}
\affiliation{Max Planck Institute for Chemical Physics of Solids, 01187 Dresden, Germany}
\author{Laurel E. Winter}
\affiliation{National High Magnetic Field Laboratory, Los Alamos National Laboratory, Los Alamos, New Mexico 87545, USA}
\affiliation{Los Alamos National Laboratory, Los Alamos, New Mexico 87545, USA}
\author{Fedor F. Balakirev}
\affiliation{National High Magnetic Field Laboratory, Los Alamos National Laboratory, Los Alamos, New Mexico 87545, USA}
\author{Ross D. McDonald}
\affiliation{National High Magnetic Field Laboratory, Los Alamos National Laboratory, Los Alamos, New Mexico 87545, USA}
\author{James G. Analytis}
\affiliation{Department of Physics, University of California, Berkeley, California 94720, USA}
\author{Nityan L. Nair}
\affiliation{Department of Physics, University of California, Berkeley, California 94720, USA}
\author{Eric D. Bauer}
\affiliation{Los Alamos National Laboratory, Los Alamos, New Mexico 87545, USA}
\author{Filip Ronning}
\affiliation{Los Alamos National Laboratory, Los Alamos, New Mexico 87545, USA}
\author{Claudia Felser}
\affiliation{Max Planck Institute for Chemical Physics of Solids, 01187 Dresden, Germany}
\author{Tobias Meng}
\affiliation{Institut fur Theoretische Physik, Technische Universit\"at Dresden, 01062 Dresden, Germany}
\author{Binghai Yan}
\affiliation{Department of Condensed Matter Physics, Weizmann Institute of Science, Rehovot, 7610001, Israel}
\author{Philip J. W. Moll}
\affiliation{Institute of Materials, \'{E}cole Polytechnique F\'{e}d\'{e}rale de Lausanne (EPFL), 1015 Lausanne, Switzerland}
\affiliation{Max Planck Institute for Chemical Physics of Solids, 01187 Dresden, Germany}


\date{\today}

\begin{abstract}
The Ce(Co,Rh,Ir)In$_5$ family of  ``Ce-115'' materials hosts an abundance of correlated electron behavior, including heavy-fermion physics, magnetism, superconductivity and nematicity. The complicated behavior of these entangled phenomena leads to a variety of exotic physical properties, which, despite the seemingly simple crystal structure of these compounds, remain poorly understood. It is generally accepted that the interplay between the itinerant and local character of Ce-$4f$ electrons is the key to their exotic behavior. Here, we report theoretical evidence that the Ce-115 materials are also topological semi-metals, with Dirac fermions around well-separated nodes. Dirac nodes in each compound are present on the $\Gamma-Z$ plane close to the Fermi level. As the Dirac bands are derived from In-orbitals, they occur in all family members irrespective of the transition metal (Co,Rh,Ir). We present the expected Fermi-arc surface state patterns and show the close proximity of a topological Lifshitz transition, which possibly explains the high field physics of Ce-115 materials. Experimentally, we highlight the surprising similarity of Ce(Co,Rh,Ir)In$_5$ in high magnetic fields, despite the distinctly different states of the Ce-$4f$ electrons. These results raise questions about the role Dirac fermions play in exotic transport behavior, and we propose this class of materials as a prime candidate for unconventional topological superconductivity.
\end{abstract}


\maketitle

\begin{figure*}[!t]
	\centering
		\includegraphics[width=0.6\linewidth]{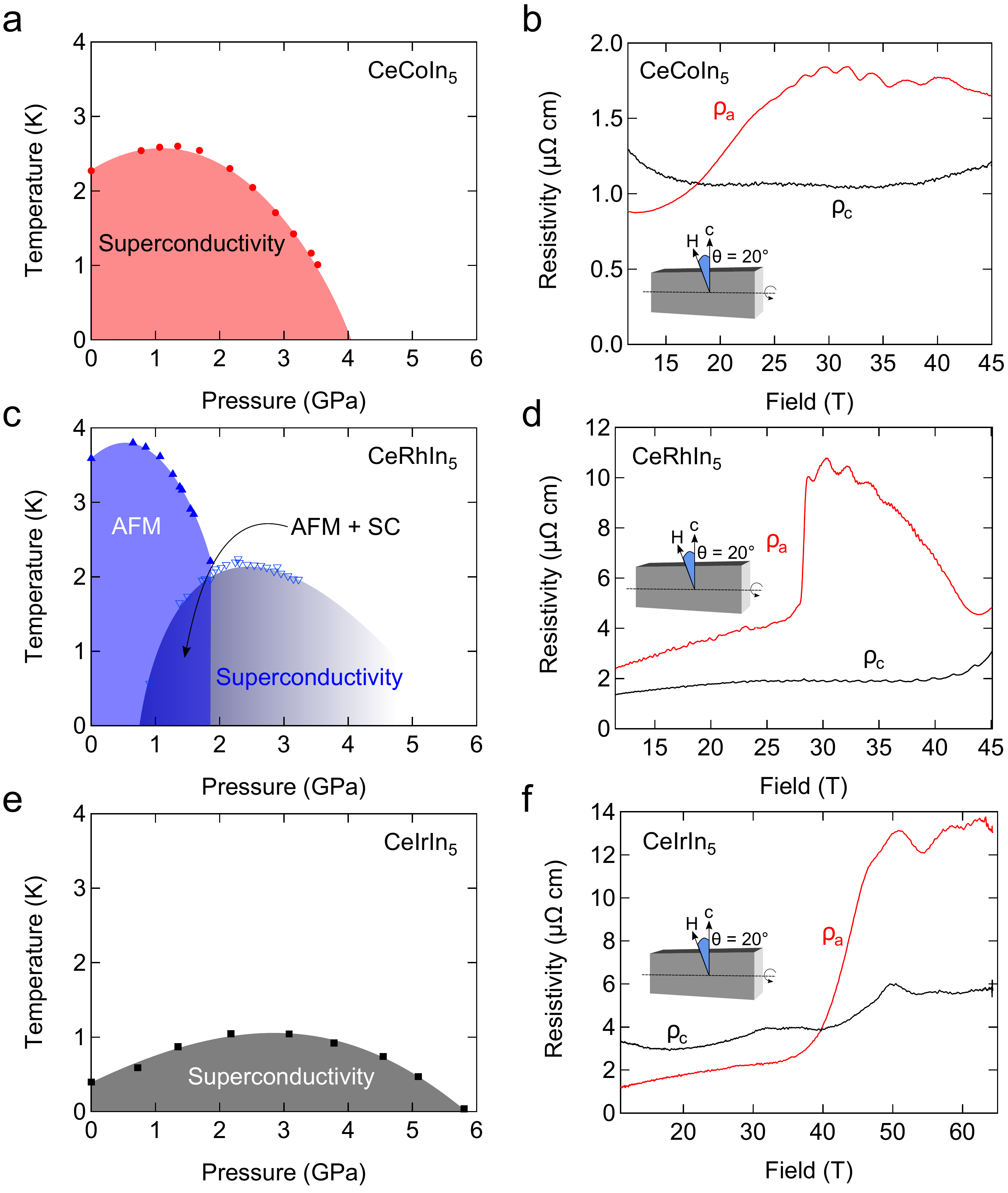}
		\caption[Phase diagrams of the Ce-115 compounds and plots of the resistivity versus field for each.]
				{
				\label{fig:Phase_Rho}
				\textbf{Phase diagrams and high field resistivity data.} Phase diagrams on the left side reproduced from Refs.~\cite{Sidorov_2002a, Knebel_2004a, Kawasaki_2006a} of the three Ce-115 compounds. Plots of the resistivity versus field are shown on the right with \CeRhIn\ data reproduced from Ref. \cite{Moll_2015a}. Despite their different behavior at zero field, as indicated by the temperature-pressure phase diagrams, the high field behavior of each material is remarkably similar. The compounds all show large quantum oscillations of similar frequency, and, in the case of \CeRhIn\ and \CeIrIn, these quantum oscillations appear after an abrupt change in the resistivity. The inset in the right side plots shows the $20^{\rm{o}}$ sample orientation with respect to the field direction.
				}
\end{figure*}

\section{Introduction} 

Intermetallic compounds containing Ce are prime examples of the rich and complex behavior which arise from strong electronic correlations. At the root of this physical behavior is the dual character of the $4f^1$ electron of Ce, which can be both local and itinerant, as well as carry a large orbital moment. This $4f$ state can hybridize with the conduction electrons, leading to non-magnetic itinerant electrons characterized by extreme quasiparticle mass enhancement, the so-called heavy fermion state. Alternatively, the $4f$ electrons can be localized, which results in local moment magnetism with often complex magnetic order \cite{Doniach_1977a}. At the boundary between phases where the dominant physics is governed by the Kondo interaction and the RKKY interaction, the hybridization tends to be very small, and perturbations such as moderate pressure, magnetic field, or chemical substitution can fundamentally change the nature of the material \cite{Loehneysen_2007}. These distinct electronic phases are separated by quantum critical points, and the quantum critical fluctuations are often quenched by the appearance of correlated electronic states, such as unconventional superconductivity. For decades, this field has significantly contributed to our knowledge of quantum criticality.

A recent advance in our understanding of solids is the realization of the importance of band structure topology, and, subsequently, the exotic properties of topological semi-metals have attracted significant theoretical and experimental efforts \cite{Wan_2011,Zyuzin_2012,Turner_2013,Xu_2014a,Sun_2015a}. These materials are characterized by band-crossings. Typically, the wavefunctions associated with the crossing points hybridize and open a gap. However, there are exceptions to this rule which arise when the symmetry of the bands prevents mixing. Such crossings are not required by crystal symmetry and were thus called ``accidental'' and generally deemed irrelevant. Yet, around these crossing points, the low energy spectrum can be linearized and the quasiparticle excitations can be mathematically mapped, depending on the band degeneracies, onto the Dirac- or Weyl-equation which governs the dynamics of massless Fermionic particles in relativistic quantum field theory \cite{Turner_2013}. The resulting quasiparticles in the solid exhibit remarkable properties originally envisioned for ultra-relativistic elementary particles. Among these exotic properties is the emergence of a concept of quasiparticle chirality. Under parallel electric- and magnetic-fields, this can be broken, which is expected to lead to an unusual negative longitudinal magnetoresistance related to the chiral anomaly in lattice models.

Topological physics is typically investigated in low carrier density, high mobility semi-metals. When the nodal points reside close to the Fermi level within the quasi-linear band dispersion, small topological Fermi surfaces can form. This is the case in the famous Dirac semimetals Cd$_3$As$_2$ and Na$_3$Bi or in the Weyl semi-metal TaAs, see Refs. \cite{Armitage_2018a,Yan_2017a} and references therein. One common misconception is that topological band structure anomalies such as nodal points are rare in nature, while, in fact, they are quite common. The seminal work of Mikitik and Sharlai has already suggested non-trivial Berry phases arising from topological defects in the band structure of LaRhIn$_5$ \cite{Mikitik_2004a}. While their quantum oscillation analysis suggests a Dirac line node which cannot be corroborated by our later results, this is a remarkable pioneering work in the field of topological matter. Accordingly, computational materials searches predict large numbers of topological materials, and many, even well studied metals, may host previously unnoticed topological nodal points in the band structure.  We propose here that the so-called 'Ce-115s' (\CeCoIn, \CeRhIn, and \CeIrIn) possess such nodal points in their band structures and show experimental and theoretical results in support of this proposal. This suggests that topological phenomena may contribute to the complex physical behavior of this materials class, and add a puzzle piece towards their understanding.

\section{Experimental Results} 

The key result reported here is the prediction of a pair of Dirac nodes in the Ce-115 band structure derived from the In states. The presence of a node is, therefore, independent of the electronic state of the Ce, as well as of the choice of transition metal (Co,Rh,Ir) and should present a common theme in this very diverse materials class. To support the commonalities of these compounds experimentally, we present resistivity data taken in high magnetic fields for all members of the Ce-115 family, shown in the right panels of Fig. \ref{fig:Phase_Rho}. Magnetoresistance measurements were performed with current applied both along the $[100]$ and $[001]$ crystal directions of magnetic fields up to $65\,\rm{T}$ for a variety of sample orientations with respect to the applied magnetic field. For clarity, we show each compound with an applied field orientation of $20^o$ from the $[001]$ toward the $[010]$ crystal direction. The angular dependent measurements were carried out on microstructured samples prepared by focused ion beam (FIB) techniques, the preparation of which is explained in detail in Ref. \cite{Ronning_2017a}. Such samples are ideal for high field experiments, especially those which utilize pulsed fields; they have low contact resistances, $~\rm{\mu m}^2$ cross-sections, and, therefore, large signal to noise.

Remarkably, all three compounds appear to behave in a very similar manner under applied magnetic field -- in particular, the large, low frequency quantum oscillations above $30\,\rm{T}$. This is surprising, as the physical difference between these materials is stark, despite their structural similarity. These differences can be summarized, in part, by the pressure-temperature phase diagrams in the left panels of Fig. \ref{fig:Phase_Rho}. \CeCoIn\ is a superconductor with a high transition temperature for a heavy fermion, 2.3 K; its electrons are itinerant, and they have a relatively large effective mass indicated by measurements of the specific heat,  $\gamma \approx 300\,\rm{mJ}/\rm{mol K}^2$ at 2.3 K and which increases dramatically at low temperatures in the presence of a magnetic field \cite{Petrovic_2001b}. In addition, \CeCoIn\ is located near a magnetic field-tuned quantum critical point (QCP) at ambient pressure. In contrast, \CeRhIn\ is an antiferromagnet, its $4f$ electron is almost entirely localized, and the effective mass of these electrons in the AFM state are the lightest of the three; the specific heat coefficient is $\gamma \approx 50\,\rm{mJ}/\rm{mol K}^2$  \cite{Shishido_2002a}. It undergoes a superconducting transition with applied pressure once the AFM order is suppressed. \CeIrIn\ is a superconductor at ambient pressure, though with a much lower transition temperature than the others, T$_c \approx 400\,\rm{mK}$. In \CeIrIn\ the effective mass of the electrons in the system are itinerant and heavy; $\gamma \approx 700\,\rm{mJ}/\rm{mol K}^2$  \cite{Petrovic_2001a}.

The application of strong magnetic fields is expected to influence these materials in different ways -- the suppression of magnetic order in the AFM compound \CeRhIn\ (which indeed occurs at even higher fields above 50T), the polarization of spin-fluctuations and, ultimately, the localization of the $4f$ electron in the cases of itinerant behavior at zero field. The experimental similarity between the Ce-115s at high fields may indicate a common thread that ties them together. A remarkable feature, shown in Fig. \ref{fig:Phase_Rho}, is the large amplitude of the quantum oscillations, the magnitude of which reaches up to $\sim10\%$ of the resistivity in \CeIrIn. The frequency of the oscillations is small, about 400 T. Clearly, the data suggest a small pocket of the Fermi surface, which contains a miniscule amount of the charge carriers, plays a significant role in the conductivity in the high field state. Additionally, \CeRhIn\ and \CeIrIn\ both experience an abrupt change in their resistivity in high fields, above which the large quantum oscillations are present.

Our hypothesis is the high field behavior in each system is dictated not only by the physics of the Ce but is also sensitive to topological physics in the In bands. The experimental evidence for this is bolstered by our band structure calculations which show the existence of a small Dirac pocket with a frequency consistent with our high field measurements.

\begin{figure*}[!t]
	\centering
		\includegraphics[width=0.95\linewidth]{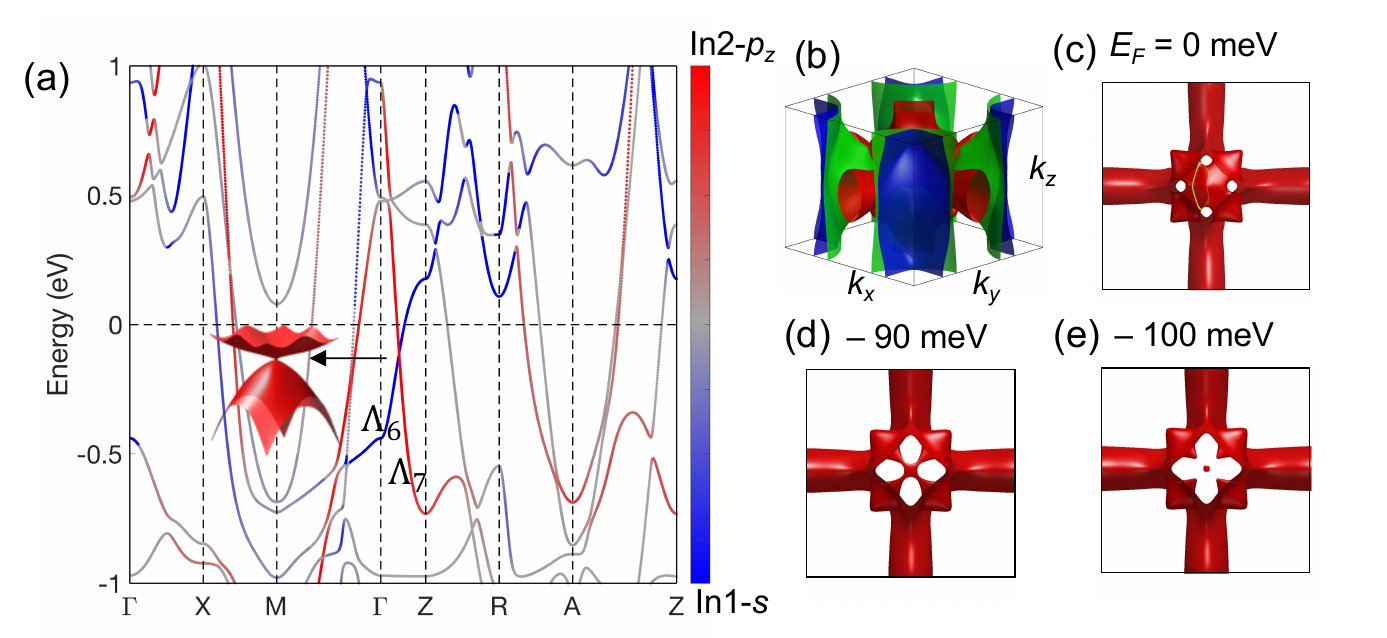}
		\caption[Electronic structure of \CeCoIn.]
				{
				\label{fig:DFT}
				\textbf{Electronic structure of \CeCoIn.} (a) The band structure. Red and blue colors represent the contribution of In2-$p_z$ and In1-$s$ orbitals, respectively. Along the $\Gamma-Z$ line, the In2-$p_z$ (band representation $\Lambda_7$) and In1-$s$ ($\Lambda_6$) cross each other, which leads to a Dirac point. The dispersion near the Dirac point in the $(k_x,k_y)$-plane is shown in inset. The Ce-$4f$ state is frozen as the core electron in the DFT calculation. The Fermi energy (E$_F$) is zero. (b) The whole Fermi surface in the Brillouin zone. Red, blue and green indicate different bands. (c)-(e) Fermi surfaces at different energies viewed from the top of the $(k_x,k_y)$-plane. For clarity, only the red part of the Fermi surface is shown. The Dirac pocket is located at the center of the zone. The green circle in (c) represents a quantum orbit caused by the Dirac pocket.
				}
\end{figure*}

\section{Bulk band structure in the density functional theory} 

We have performed band structure calculations for each of the La- and Ce-115 compounds using density-functional theory (DFT) within the generalized-gradient approximation. To treat the pseudopotential of Ce, we froze the Ce-$4f$ as the core electron, so that both La and Ce exhibit three valence electrons. All materials possess an intriguing common feature, a Dirac point slightly below the Fermi energy in their band structure. Here, we discuss the features of  these band structure calculations common to all the Ce-115s, using \CeCoIn\ as an example. At nearly the midpoint of the $\Gamma-Z$ line, there is an accidental band crossing at about $110\,\rm{meV}$ below the Fermi energy (E$_F$). Because all bands are doubly degenerate, caused by the coexistence of time-reversal symmetry (TRS) and inversion symmetry, such a band crossing point is four-fold degenerate, which results in a Dirac point. The Dirac point is protected by the $C_4$ rotational symmetry and comes in pairs (below and above the $\Gamma$ point), similar to the known Dirac semimetal Cd$_3$As$_2$ \cite{Wang_2013b}. In other words, such a band crossing is unavoidable, as different band representations forbid the hybridization between them. We note that $\Lambda_6$ and $\Lambda_7$, labeled in Fig. \ref{fig:DFT}, are double-group representations of the space group $P4/mmm$ (No.123).

By analyzing the orbital contribution, we find that all Dirac bands are mainly contributed by In states. There are two groups of In atoms in the tetragonal unit cell of \CeCoIn. Four In are located on the side surfaces (indicated as In2) and one sits on the top surface (In1). The In1-$s$ and In2-$p_z$ orbitals contribute to $\Lambda_6$ and $\Lambda_7$ bands, respectively. Along the $\Gamma-Z$ direction, In1-$s$ bands disperse up while In2-$p_z$ bands disperse down, due to different orbital symmetry between them. As In sites are common to the Ce-115s, it is not surprising to observe that each compound exhibits the same In-state-induced Dirac point in the band structure (see the supplementary information for details of all compounds). It is known that DFT usually fails to describe the $d$- or $f$-states of the correlated electron system. However, the Dirac point is expected to remain robust in the 115 materials, since it is composed by the weakly-interacting $sp$ states. We point out that such a band crossing point can also be found in the earlier theoretical work  \cite{Choi_2012a}, although its topological aspect was not appreciated.  In Ref. \cite{Choi_2012a}, dynamic mean field theory revealed that these Dirac bands may be slightly renormalized when coexisting with $d$- and $f$-states while the Dirac point remains intact.

The whole Fermi surface of \CeCoIn\ is large and complicated. In contrast, Dirac bands induce a tiny and anisotropic Fermi surface that only corresponds to nearly $1\%$ of the total carrier density. When E$_F$ is close to the Dirac point, the Dirac pocket is an isolated, flattened sphere. As E$_F$ increases away from the Dirac point, the sphere grows and its boundaries start to merge into another larger Fermi surface -- a Lifshitz transition, as shown in Fig. \ref{fig:DFT}. At E$_F$ $= 0$, the Dirac pocket is partially connected to a large Fermi surface sheet. When greater than 30 degrees off the $k_z$-axis, a closed quantum orbit emerges from the Dirac Fermi surface (see Fig. \ref{fig:DFT}c).

\section{Topological surface states}

\begin{figure*}
	\centering
		\includegraphics[width=0.8\linewidth]{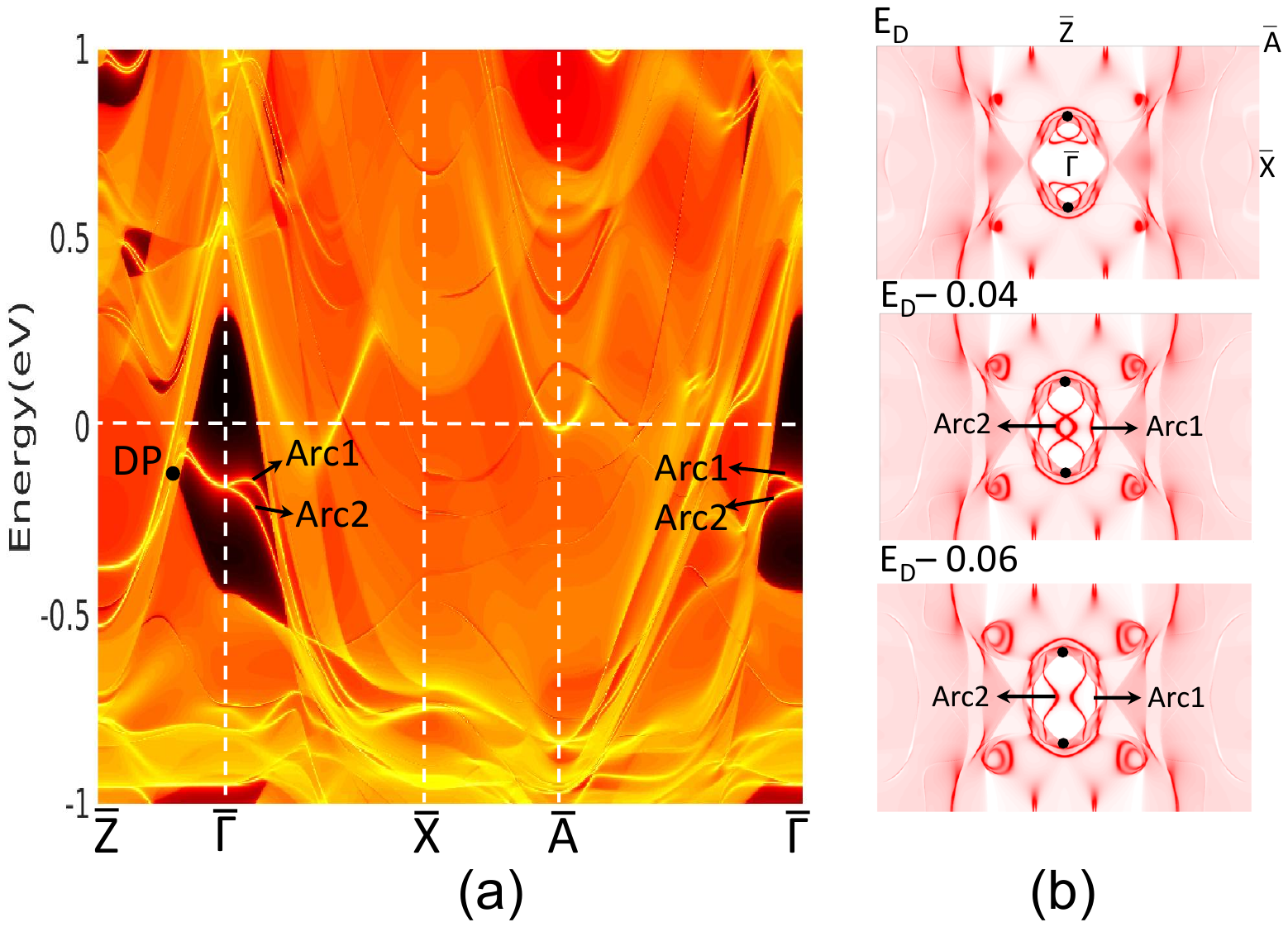}
		\caption[Surface states with Fermi arcs]
				{
				\label{fig:FermiArcs}
				\textbf{The surface state with Fermi arcs.} (a) The surface band structure of a half-infinite $(x,z)$-plane. Two surface states (labeled Arc1 and Arc2) merge at the Dirac point (DP, the black dot). The Femi energy is set to zero. (b) The constant energy surface for energies at the Dirac point ($E_D$),  $E_D - 0.04$, and $E_D - 0.06\,\rm{eV}$. The black dots indicate the DP positions. 
				}
\end{figure*}

Topological surface states appear on the side surface of a Dirac metal because of a bulk-boundary correspondence. As shown in Fig. \ref{fig:FermiArcs}, two surface bands appear inside the local bulk energy gap at the $\overline{\Gamma}$ point as a Kramers pair, due to TRS, and merge together at the projection of bulk Dirac points between $\overline{\Gamma}$ and $\overline{Z}$. This gives rise to the topological Fermi arc states on the $xz$ plane of the crystal, and two surface bands become bulk resonances at large momenta along $\overline{\Gamma}-\overline{X}$ and $\overline{\Gamma}-\overline{A}$. The corresponding Fermi surface, the constant energy contour of the band structure, exhibits peculiar features. For an ordinary Dirac semimetal (e.g. Na$_3$Bi) \cite{Xu_2014a}, the upper and lower surface bands are hole-like and electron-like in dispersion, respectively. Consequently, the Fermi surface contour through the Dirac point only crosses the upper or lower bands and results in two simple Fermi arcs that link two Dirac points. Here, in contrast, both upper and lower surface bands of \CeCoIn\ are hole-like and display strong anisotropy in their dispersion. Therefore, the constant energy surface at the Dirac point intersects both the upper and lower surface bands, denoted Arc1 and Arc2 respectively. This induces two ring-like loops (see Fig. \ref{fig:FermiArcs}b). Below the Dirac point energy, both Arc1 and Arc2 emerge to link the two Dirac points. In reality, the specific shape and connectivity of these Fermi arcs are also sensitive to the surface conditions, similar to those of the Weyl semimetal TaAs \cite{Sun_2015a}. We stress that the topology of the current surface states is fully equivalent to an ordinary Dirac semimetal, which is characterized by the existence of a single pair of helical states originating from the Dirac points. These Fermi arc states, which lie below the Fermi energy, provide a hallmark to identify bulk Dirac points in future surface spectroscopy measurements such as angle-resolved photoemission spectroscopy.

\section{Discussion} 

Undoubtedly, a key ingredient to the complex physics of Ce-115 materials are the strongly interacting Ce-$4f$ electrons. It is a priori unclear if the Dirac nodes we have identified when the $4f$ electrons are treated as core electrons survive when interactions are addressed more comprehensively: the impact of electron-electron interactions on topological features is, in general, unknown, and remains a vibrant field of research.

For three-dimensional Weyl and Dirac semimetals, there is corroborating evidence that their nodal points are rather robust to interaction effects  \cite{Isobe_2012,Isobe_2013,Sekine_2014,Xiao_2018}. Power counting, for example, shows that weak short-ranged interactions are irrelevant in the renormalization group (RG) sense at linearly dispersing nodal points. Weak long-range interactions, on the other hand, are marginally irrelevant, and hence give rise to logarithmic corrections to physical properties \cite{Abrikosov_1971,Goswami_2011,Isobe_2012,Hosur_2012}. It has, furthermore, been shown that nodal points are perturbatively stable to all orders \cite{Bergholtz_2018,Carlstroem_2018}.

This does not mean, however, that interactions cannot lead to intriguing physics. On the one hand, sufficiently strong interactions can stabilze entirely new phases, including axion insulators \cite{Wan_2011,Zyuzin_2012,Go_2012,Wang_2013a}, orthogonal semimetals in which electrons fractionalize into slave-fermions, and a $\mathds{Z}_2$-spin \cite{Witczak_2014}; even more exotic metals exhibit a fractional variant of the chiral anomaly \cite{Meng_2016}, as well as semimetals with single unpaired Weyl nodes \cite{Meng_2018} or a Mott gap \cite{Morimoto_2016}. On the other hand, different kinds of interactions can also drive the formation of three-dimensional topological semimetals \cite{Wan_2011,Go_2012,Savary_2014,Moeller_2017}.

The specific combination of heavy fermion physics and nodal semimetals in Kondo lattices was studied in Ref.~\cite{Feng_2016} for the case of a two-dimensional Dirac semimetal. It was found that Dirac nodes can remain stable both at weak and strong interactions on the $4f$ sites. Depending on the correlations, however, the band structure may be vastly renormalized. In three-dimensional Kondo lattices, it has similarly been shown that the interacting $4f$ electrons can drive the formation of Weyl-Kondo semimetals \cite{Lai_2017} in which the electrons at the Weyl node inherit the $4f$ electron's heaviness. Experimentally, their strongly reduced velocity, for example, implies a particularly large value of the specific heat.

The general robustness of topological semimetals with respect to interactions suggests that Dirac physics may survive in Ce-115 compounds even if the strongly correlated $4f$ electron are not treated as core electrons. Intriguingly, we can retrieve the Dirac nodes present in our numerics in earlier band structure calculations of these well-studied heavy fermion compounds -- they have simply been overlooked until now. The results of Elgazzar \textit{et al.} \cite{Elgazzar_2004} based on the full potential local orbital (FPLO) method, for example, reveal that when the $4f$ electrons are not treated as core electrons, the most substantial modification of the band structure is the appearance of rather flat bands around the Fermi energy. Importantly, one can still discern a band crossing a few hundred meV below the Fermi level on the $\Gamma-Z$ line. Similarly, a nodal point is apparent around $250\,\rm{meV}$ below the Fermi level in the closely connected family of compounds Pr(Co,Rh,Ir)In$_5$, in which Pr replaces Ce as the heavy element which introduces strong correlations \cite{Elgazzar_2008}. The persistence of the Dirac node is also supported by the Generalized Gradient Approximation method (GGA and GGA+U) in Ref.~\cite{Wang_2003}, and Density-Functional Theory combined with Dynamical Mean-Field Theory (DFT+DMFT) as presented in the Supplemental Material of Ref.~\cite{Choi_2012a}: also these studies feature a crossing in their band structures that we identify with the Dirac node of our calculations.

These calculations indicate that the orbital character of the bands which form the Dirac node changes as a function of momentum in the vicinity of the node. This varying orbital character also seems to persist when the $4f$ electrons are not treated as core electrons. In Ref.~\cite{Choi_2012a}, we observe correlation effects to have different strengths on different parts of the bands which form the nodal point  -- an observation that can naturally be explained if the orbital content is momentum-dependent. The change of orbital character is also present in the band structure of Ref.~\cite{Elgazzar_2008}.

In light of these earlier numerical results, we propose that a minimal theoretical model for Ce-115 compounds can be obtained by the hybridization of a flat Ce band close to the Fermi level with the band structure of the respective La-115 compound. To properly account for the momentum-dependence of correlation effects, the hybridization should reflect the orbital character of the respective bands. Technically, the hybridization between the $4f$ electrons and the La-115 band structures may be modeled using the the slave-boson formalism \cite{Barnes_1976,Barnes_1977,Coleman_1984,Kotliar_1986,Hewson_book}. The latter starts from a periodic Anderson model of conduction electrons coupled to strongly interacting $4f$ electrons. The corresponding Hamiltonian can be decomposed as $H=H_c+H_f+H_{cf}$ with

\begin{align}
H_c &= \sum_{{\mathbf{k}},\sigma,\alpha}(\epsilon_{{\mathbf{k}}\sigma \alpha}-\mu)\,c^\dagger_{{\mathbf{k}}\sigma \alpha} c^\pd_{{\mathbf{k}}\sigma \alpha},\\
H_f &=\sum_{j,\sigma} (E_{f\sigma}-\mu)\,f_{j\sigma}^\dagger f_{j\sigma}^\pd +\sum_{j}U\,f_{j\uparrow}^\dagger f_{j\uparrow}^\pd f_{j\downarrow}^\dagger f_{j\downarrow}^\pd,\\
H_{cf}&=\sum_{\mathbf{k},\sigma,\alpha,j} V_{\mathbf{k}\alpha}\,e^{i \mathbf{k}\cdot{\mathbf{R}}_j}\,c_{\mathbf{k}\sigma \alpha}^\dagger f_{j\sigma}^\pd+\rm{H.c.},
\end{align}
where $c_{\mathbf{k}\sigma \alpha}$ is the annihilation operator for conduction electrons of momentum $\mathbf{k}$ and spin $\sigma$ in band $\alpha$, and $f$-electrons on site $j$ located at $\mathbf{R}_j$ are annihilated by $f_{j \sigma}$. For spin $\sigma$, the energy of band $j$ is $\epsilon_{{\mathbf{k}}\sigma j}$, while the energy of the $f$-electrons is $E_{f\sigma}$. The on-site interaction reads $U$, $\mu$ is the chemical potential, and the $V_{\mathbf{k}\alpha}$ denotes the hybridization between the the $f$-electrons and band $\alpha$.

In the limit of infinite interactions between the $4f$ electrons, $U\to\infty$, double occupancy of the $f$-sites is forbidden. One can then decompose the $f$-electrons into auxiliary fermions $\tilde{f}$ and slave-bosons $e$ representing empty sites as $f_{j \sigma}^\pd=e_j^\dagger \tilde{f}_{j\sigma}^\pd$ subject to the constraint $e_j^\dagger e_j^\pd+ \sum_\sigma \tilde{f}_{j\sigma}^\dagger \tilde{f}_{j\sigma}^\pd=1$. Introducing Lagrange multiplier $\lambda_j$ for this constraint yields the slave-boson Hamiltonian

\begin{align}
H_{\rm SB} &= \sum_{{\mathbf{k}},\sigma,\alpha}(\epsilon_{{\mathbf{k}}\sigma \alpha}-\mu)\,c^\dagger_{{\mathbf{k}}\sigma \alpha} c^\pd_{{\mathbf{k}}\sigma \alpha},\nonumber\\
&+\sum_{j,\sigma} (E_{f\sigma}-\mu+\lambda_j)\,\tilde{f}_{j\sigma}^\dagger \tilde{f}_{j\sigma}^\pd+\sum_j \lambda_j \left(e_j^\dagger e_j^\pd-1\right)\nonumber\\
&+\sum_{\mathbf{k},\sigma,\alpha,j} \left(V_{\mathbf{k}\alpha}\,e^{i \mathbf{k}\cdot{\mathbf{R}}_j}\,c_{\mathbf{k}\sigma \alpha}^\dagger e_j^\dagger \tilde{f}_{j\sigma}^\pd+\rm{H.c.}\right).
\end{align}
In its simplest variant, the slave-boson mean-field approximation assumes that the bosonic fields and Lagrange multiplier take constant values, $e_j\to e$ and $\lambda_j \to \lambda$. The latter can be chosen as real numbers since the phase of the mean-fields can be absorbed into $\tilde{f}_{j\sigma}$ via a gauge transformation. The mean-field parameters are determined by minimization of the free energy with respect to $e$ and $\lambda$. This yields the mean-field equations

\begin{align}
e^2&=1-\frac{1}{\mathcal{N}}\sum_{j,\sigma} \langle \tilde{f}_{j\sigma}^\dagger \tilde{f}_{j\sigma}^\pd\rangle,\\
2\lambda e &=-\frac{1}{\mathcal{N}}\sum_{\mathbf{k},\sigma,\alpha,j} V_{\mathbf{k}\alpha}\,e^{i \mathbf{k}\cdot{\mathbf{R}}_j}\,\langle c_{\mathbf{k}\sigma \alpha}^\dagger \tilde{f}_{j\sigma}^\pd \rangle+\rm{H.c.},
\end{align}
where $\mathcal{N}$ is the number of $4f$-sites. In addition, the chemical potential has to be adjusted to keep the total particle number $N_{\rm tot}$ fixed, $N_{\rm tot} = \sum_{{\mathbf{k}},\sigma,\alpha}\langle c^\dagger_{{\mathbf{k}}\sigma \alpha} c^\pd_{{\mathbf{k}}\sigma \alpha}\rangle + \sum_{j,\sigma} \langle \tilde{f}_{j\sigma}^\dagger \tilde{f}_{j\sigma}^\pd\rangle\stackrel{!}{=}N_c+N_f$, where $N_{c}$ ($N_f$) is the numbers of the original $c$ ($f$) electrons. Upon calculating the Green's function of the original fermions, one finds that their quasiparticle weight is determined by the mean-field parameter $|e|^2$.

The slave-boson mean field approach can be refined in a number of ways. Finite interactions $U$ can for example be taken into account by mean-field schemes that also allow double occupancies of the $f$-sites \cite{Kotliar_1986}. To more accurately describe the momentum dependence of correlation effects, one may also retain an explicit momentum-dependence of the mean-field parameters. A detailed slave-boson study based on our numerical results will be presented elsewhere.

\section{Conclusion} 

We have presented new experimental and theoretical evidence for the existence of topologically protected states within a heavy fermion framework in the Ce-115 materials, with Dirac nodes located on the $\Gamma-Z$ plane close to the Fermi level. We showed how the Dirac bands are derived from In-orbitals, and these bands occur in all family members irrespective of the transition metal (Co,Rh,Ir). The presence of such Dirac bands in the Ce-115s may indeed provide insight into some of the outstanding mysteries in this class of heavy fermion materials.

The DFT calculations presented here provide an initial step towards our understanding of the In-derived bands. However, a more realistic treatment of the $4f$-electrons is an essential ingredient in determining the fate of the Dirac pocket in the presence of correlations and, potentially, magnetism. We must, therefore, turn to the unresolved debate about the importance of topological aspects in metals. On one hand, accidental band degeneracies are commonplace in many metals; the physical distinction between linear and quadratic dispersions at half band filling may appear marginal, and the number of topological charge carriers comprise only a few percent of the total density of states. So, at first glance, the presence of a small Dirac Fermi surface may not be relevant. On the other hand, the presence of topological charge carriers may impact correlated states more strongly than their number suggests. The non-trivial spin-momentum-locking and Berry curvature of electrons at a Dirac node for example necessarily affects the superconducting pairing, which possibly becomes topological \cite{hosur_sc_14,kobayashi_15,hashimoto_16,chirolli_17,haldane_18}. This scenario has by now been substantiated by experimental reports in favor of unconventional superconductivity in Dirac semimetals \cite{aggarwal_16,he_16,oudah_16,wang_16,wang_17}. The tendency towards unconventional pairing may be further strengthened by magnetic impurities \cite{rosenstein_15a,rosenstein_15b}. 

\begin{figure}
	\centering
		\includegraphics[width=0.8\linewidth]{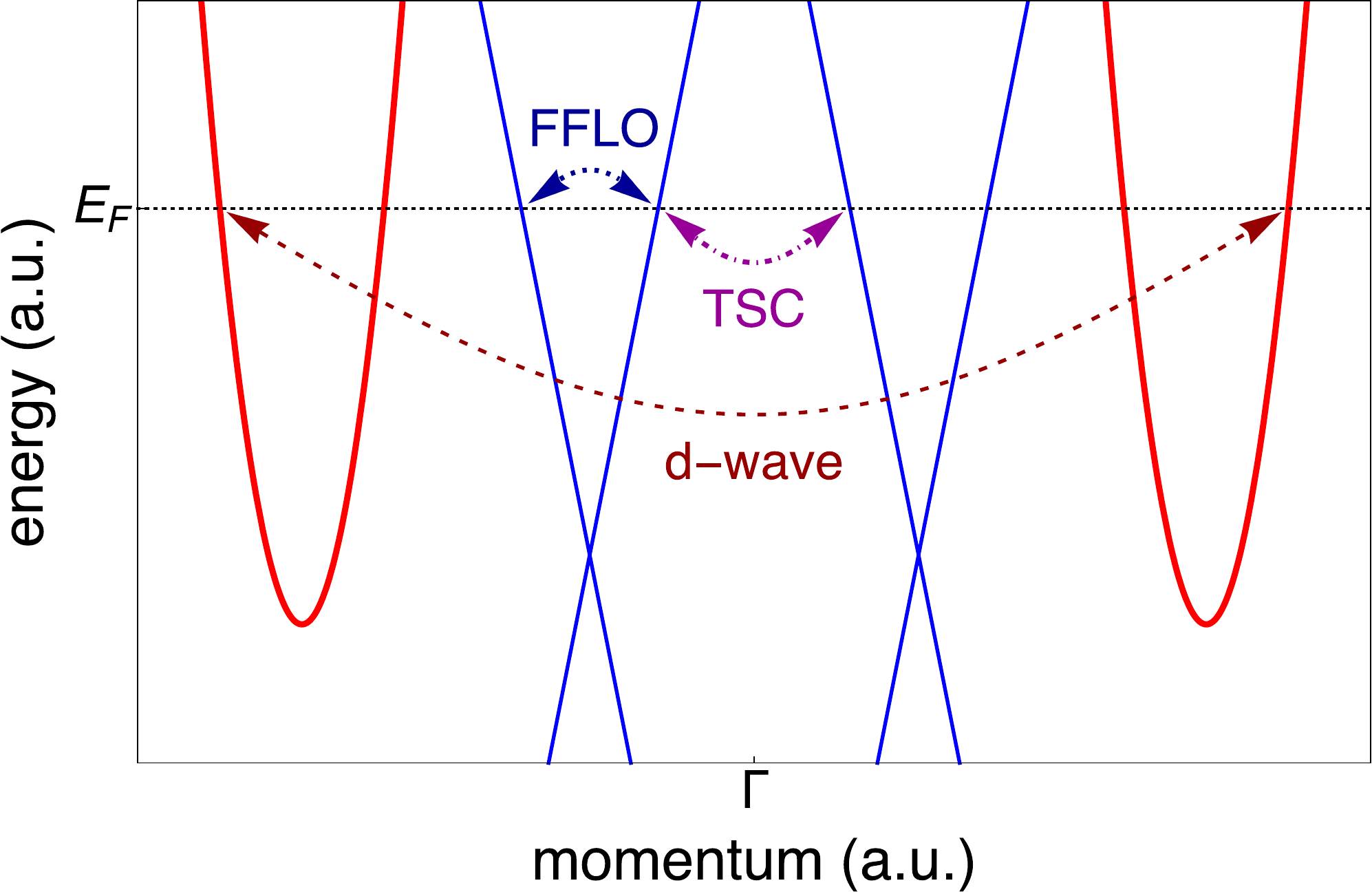}
		\caption[Surface states with Fermi arcs]
				{
				\label{fig:sketch_SC}
				\textbf{Possible superconducting states in {\CeCoIn}.} As explained in the main text, Dirac bands (blue) close to the Fermi energy $E_F$ have a tendency towards unconventional superconducting states, such as topological superconductivity (TSC) or pairing with a finite momentum (FFLO). Additionally present trivial bands (red) are experimentally found to develop $d$-wave superconductivity. 
				}
\end{figure}

Besides topological pairing with Cooper pairs of zero momentum, it has also been argued that topological semimetals allow for intranode pairing with a finite momentum, i.e.~a Fulde-Ferrell-Larkin-Ovchinnikov (FFLO) state \cite{ff_64,lo_65,cho_12,aji_14,bednik_15,zhou_16}. As illustrated in Fig.~\ref{fig:sketch_SC}, however, any kind of unconventional superconductivity at the Dirac nodes would have to compete with more standard superconducting orders in other, non-topological bands at the Fermi surface. As a result, the Dirac pockets may form a topological superconducting state that does not extend to the trivial Fermi surfaces in \CeIrIn. Conversely, superconducting order in the trivial Fermi pockets may not spread to the Dirac pockets, or only with a much reduced gap. Remarkably, measurements of the specific heat and thermal conductivity at low temperatures in \CeCoIn\ revealed that a small portion of the conduction electrons does not participate in superconductivity. This result suggested an exotic case of multiband superconductivity, where two groups of electrons are seemingly decoupled from one another \cite{Tanatar_2005a}. Later work has suggested a very small but finite gap on a small Fermi surface. It is interesting to note that the total fraction of Dirac electrons estimated by DFT is around 2\% (for the twofold degenerate Fermi surface) and is close to the reported unpaired fraction of 3\%. Similar arguments may apply to the formation of the heavy fermion state itself. A two-fluid analysis suggests that less than 10\% of the $f$-electrons remain unhybridized in the ground state \cite{Nakatsuji_2004a}.

Additionally, the high field transition in \CeRhIn\, signaled by a discontinuous jump of the resistivity at $H^*\approx 28T$, has recently been suggested to be of nematic character. While this transition changes the in-plane conductivity by a factor of 10, the complete absence of discernible features in the magnetization \cite{Shishido_2002a} and the magnetic torque \cite{Moll_2015a} at $H^*$ clearly suggests an electronic transition in absence of metamagnetism. A similar transition occurs in \CeIrIn\ (Fig. \ref{fig:Phase_Rho}), despite the fact there is no magnetic order to begin with  \cite{Aoki_2016a}. These observations strongly indicate a high field transition in the electronic system that is independent of the Ce-$4f$ electron. The Dirac node, however, provides a natural origin of this transition and suggests it to be a topological Lifshitz transition. The Dirac pocket is connected to a topologically trivial Fermi sheet via thin bridges along the [110] direction (Fig. \ref{fig:DFT}d,e). Under strong magnetic fields, an energetically proximate Lifshitz transition would occur with the detachment of the Dirac pocket. Such a transition naturally has strong influence on the magnetoconductance as it changes the nature of orbits from open to closed. The appearance of giant quantum oscillations in this situation is strongly reminiscent of that in elemental Al, where a breakdown orbit connecting larger Fermi sheets via a small pocket leads to an unusual enhancement of the Shubnikov-de Haas amplitude \cite{Lonzarich_1985a}. This simple picture provides a natural origin for the XY-nematicity, where, under tilted applied magnetic fields, the bridges may be broken selectively leading to sizable in-plane anisotropy of the conductivity.

Given that the Dirac point is a consequence of In1-$s$-states and In2-$p_z$-states, we would anticipate that similar physics may occur in other related systems such as CeIn$_3$, Ce$_2$RhIn$_8$, CePt$_2$In$_7$, as well as other heavy fermion systems more broadly.

Finally, it is worthwhile to discuss an interesting property of topological band structure defects, where a conservation law for quasiparticle chirality emerges on the associated Fermi surfaces. In the present case, a massless Dirac fermion may be considered as a superposition of two Weyl fermions, one left- and one right-handed. This may be relevant to the magnetism in \CeRhIn. Magnetic frustration leads to a chiral magnetic order within the inversion symmetric crystal structure. Neutron scattering experiments have made the remarkable observation that even large scale crystals can be in a single chiral domain \cite{Fobes_2017}. Other groups have reported chiral domains, though with a remarkably high preference for one chirality. A macroscopic monochiral state in a crystal is surprising given the subtle energetic balance necessary for the emergence of an incommensurate spin-spiral ($T_N\approx 3.85K$). The occurrence of chiral magnetism along the crystallographic c-direction is allowed by symmetry to generate an imbalance between the quasiparticle chiralities, thus splitting the Dirac Fermi surface into Weyl surfaces of different size. The chirality imprinted into the itinerant electronic system may provide a long-ranged interaction, which would allow a frustrated local-moment magnet to establish a macroscopic mono-domain chiral state. The Ce-115 family provides an ideal testing ground to explore these ideas and to explore the coexistence of strong correlations and a Dirac Fermi-surface in very clean systems.

\begin{acknowledgments} 

T.M. is supported by the Deutsche Forschungsgemeinschaft via the Emmy Noether Programme ME 4844/1-1 and through SFB 1143. B.Y. is supported by a research grant from the Benoziyo Endowment Fund for the Advancement of Science and by the collaborative Max Planck Lab on ``Topological Materials''. F.R. and E.D.B acknowledge funding from the Los Alamos National Laboratory LDRD program. Work at the National High Magnetic Field Laboratory was supported by National Science Foundation Cooperative Grants DMR-1157490 and DMR-1644779, the US DOE, and the State of Florida.

\end{acknowledgments}

\section{Supplementary Information}

We have included band structure calculations for the La- and Ce-115 compounds. For each compound a Dirac-type crossing is present.

\bibliography{2018.04_TopologicalHeavyFermions_refs}

\renewcommand{\thefigure}{S\arabic{figure}}

\setcounter{figure}{0}

\begin{figure*}[!h]
	\centering
		\includegraphics[width=\linewidth]{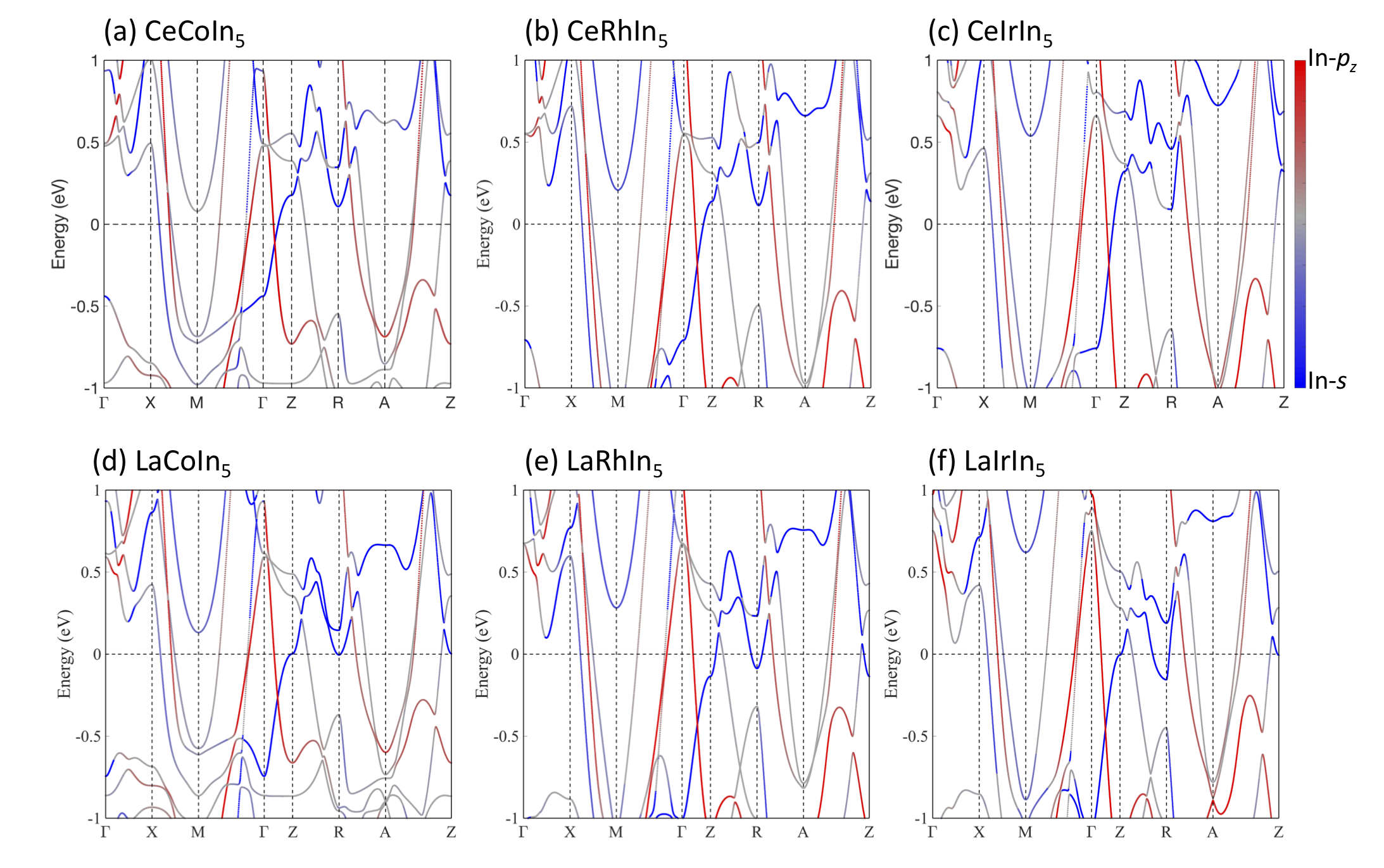}
		\caption[Band structure of 115s]
				{
				\label{fig:BS_115s}
				\textbf{Band structure of the 115 compounds.} Band structure calculations are shown for each Ce-115 compound and its non-magnetic analogue containing La. The red and blue colors represent the contribution of In-$p_z$ and -s orbitals, respectively. Along the $\Gamma-Z$ line, a Dirac-type band crossing appears slightly below the Fermi energy (zero) for all compounds.
				}
\end{figure*}

\end{document}